\documentclass[12pt]{article}
\usepackage{graphicx}
\begin {document}
\title {\bf Reduced quantum actions for symmetrical and some separated variables potentials cases.}
\author{ T.~Djama\footnote{Electronic address:
{\tt djamatoufik@yahoo.fr}}}
\date{\today}
\maketitle
\begin{abstract}
\noindent In this paper, we pointed out the separability of the
quantum reduced action in 3D into the sum of three
1D reduced actions depending on the variables
$x$, $y$ and $z$ respectively, and this was done for
the case of a potential that has a cartesian symmetry.
This separability was not evident at the first sight.
In addition, the 3D-QSHJE is also separable into
three 1D-QSHJE. The free particle, the spherical
and the cylindrical symmetry cases are not discussed in this paper,
however an analogy with the cartesian symmetry case
can be down to show the separability of the total reduced actions
into the sum of three 1D reduced actions for each case.
\end{abstract}
\vskip\baselineskip \noindent PACS: 03.65.-w

\newpage

In the last decade, in the same school of trajectory representation
\cite{Floyd1, Floyd2, Floyd3, Floyd4, Floyd5,Farmat1, Farmat2, Farmat3}
a new approach of quantum mechanics
has raised from our works \cite{Dja1, Dja2, Dja3} and developed
in order to construct a deterministic dynamical
approach of quantum mechanics.
We started to build a consistent one dimensional theory,
then we generalized our approach to 3D systems in earlier
works \cite{Dja4, Dja5, Dja6}. As an application of our theory,
we studied quantum trajectories of the Hydrogen atom's electron. For this case,
the symmetry of the potential is spherical and we proposed to separate
the total reduced action into the sum of three 1D actions \cite{Dja4, Dja5, Dja6}.
However, we have been criticized by
A. Bouda \cite{boud1,boud2} in our approach of the
symmetrical potential problems. In fact we considered mainly that
for these cases, the 3D reduced action can be written as the sum of
three 1D reduced actions each one depending on one variable. For example
for the potentials that has a cartesian symmetry
\begin{equation}\label{potsep}
V(\vec{ r})\, =\, V_x(x)+V_y(y)+V_z(z) \; ,
\end{equation}
a possible solution for the
reduced action is
\begin{equation}\label{sep-action}
S_0^{(1)}(\vec{r})= S_{0x}(x)+S_{0y}(y)+S_{0z}(z) \; ,
\end{equation}
where
\begin{eqnarray}\label{1d-action}
S_{0x}(x)=\hbar \arctan{\left({X_1+\gamma_{1}X_2\over
\gamma_{2}X_1+X_2}\right)}\, ,\nonumber\\ S_{0y}(y)=\hbar
\arctan{\left({Y_1+\gamma_{3}Y_2\over \gamma_{4}Y_1+Y_2}\right)}\,
,\nonumber\\ S_{0z}(z)=\hbar \arctan{\left({Z_1+\gamma_{5}Z_2\over
\gamma_{6}Z_1+Z_2}\right)}\, ,\nonumber
\end{eqnarray}
up to an additive constant, are solutions of the
1D-QSHJE with respect to $x$, $y$ and $z$. $X_1$ and $X_2$ are two independent solutions of the 1D Schr\"odinger equation (SE) with respect to $x$. $Y_1$ and $Y_2$ are two independent solutions of the 1D SE with respect to $y$. $Z_1$ and $Z_2$ are two independent solutions of the 1D SE with respect to $z$.
Another example is the free particle case considered in 3D. For the free particle, the general solution of the SE has the form
\begin{equation}\label{Psifreepart}
\Psi_{lm}(r,\theta,\phi)=\sum_{i, j, k=1}^2 \alpha_{ijk}\, U_i^{l}(r) T_j^{lm}(\theta) F_k^{m}(\phi)
\end{equation}
and a solution for the reduced action is
$$
S_{0_{lm}}^{(1)}(\vec{r})= S_{0r}^{l}(r)+S_{0\theta}^{lm}(\theta)+S_{0\phi}^{m}(\phi) \; ,
$$
where
\begin{eqnarray}
S_{0r}^{l}(r)=\hbar \arctan{\left({U_1^{l}+\eta_{1}U_2^{l}\over
\eta_{2}U_1^{l}+U_2^{l}}\right)}\, ,\nonumber\\
S_{0\theta}^{lm}(\theta)=\hbar
\arctan{\left({T_1^{lm}+\eta_{3}T_2^{lm}\over \eta_{4}T_1^{lm}+T_2^{lm}}\right)}\,
,\nonumber\\
S_{0\phi}^{m}(\phi)=\hbar \arctan{\left({F_1^{m}+\eta_{5}F_2^{m}\over
\eta_{6}F_1^{m}+F_2^{m}}\right)}\, ,\nonumber
\end{eqnarray}
are solutions of the 1D-QSHJE with respect to $r$, $\theta$ and $\phi$.
$U_1^{l}=r^l$ and $U_2^{l}=r^{-l-1}$ are two
independent solutions of the 1D SE with respect to $r$.
$T_1^{lm}=\sqrt{{2l+1\over 2}{(l-m)!\over (l+m)!}}P_l^m(cos\theta)$, where $P_l^m(cos\theta)$
is the Legendre associated polynomial, is the solution of the 1D SE written with respect
to $\theta$, and $T_2^{lm}$ is its corresponding independent solution.
$F_1^{m}=\sin(m\phi)$ and $F_2^{m}=\cos(m\phi)$.
Here, it is useful to emphasis that the free particle case is an interesting one,
however, in this paper, we will just discuss the case where the potential is separable
and has the form given in Eq. (\ref{potsep}). For, the free particle case, the writing of the 3D
reduced action as the sum of three 1D reduced actions can be shown in the same way
as for the separable potential given in Eq. (\ref{potsep}).

For the separable potential case (Eq. (\ref{potsep})), a solution of the stationary SE is
$$
\psi(x,y,z)=\phi_{x}(x)\phi_{y}(y)\phi_{z}(z) \; ,
$$
and the 3D-SE separates into three 1D equations similar to the
1D-SE\footnote{This not the case for some separated variables
potentials. For example, a potential of the form $$V(\vec{r}) =
V_r(r)+V_{\theta}(\theta)+V_{\phi}(\phi)\, ,$$ the SE written with spherical coordinates does not separate into three equations of one variable, simply because $V(r,\theta,\phi)$ do not hold the symmetry of the Laplacian written with spherical coordinates. Meanwhile, For all symmetrical potentials, the separation in the SE can be done. Some other potentials cannot be separated into 1D potentials that depend on one variable, but hold the SE symmetry that allows the separation in the SE, an example is the potential of the form $$V(\vec{r}) =
V_r(r)+{V_{\theta}(\theta)\over r^2}\, .$$ This why, through this article, I use more the adjective "symmetrical" than "separated" or "separable" when talking about potentials.}
\begin{eqnarray}\label{1d-schro-x}
-{\hbar^2 \over 2m}{d^2\phi_{x_i} \over
dx_i^2}+V_{x_i}(x_i)\phi_{x_i}(x_i)=E_{x_i}\phi_{x_i}(x_i) \; ,
\\ i=1,2,3.\hskip8mm  x_1=x, x_2=y, x_3=z\hskip5mm\nonumber
\end{eqnarray}
In the following
we consider the case when the wave function has the form
\begin{equation}\label{psitot1}
\Psi(x,y,z)=\sum_{i, j, k=1}^2 a_{ijk}\,  X_i(x) Y_j(y) Z_k(z) \;
,
\end{equation}
where $a_{ijk}$ are real constants. Here, note that this
last solution is not the more general one, of course for
the free particle a more general solution is the one given
in Eq. (\ref{Psifreepart}), however, as stated above, we will restrict ourselves
to the solution of the form (\ref{psitot1}).

\noindent Notice that expression (\ref{psitot1}) is also a solution
of each of the three Eqs. (\ref{1d-schro-x}). In the literature of the
Bohmian mechanics and the trajectory approaches of
quantum mechanics, the wave function $\Psi(x,y,z)$ is given as
\begin{equation}\label{wavefunction1}
\Psi(x,y,z)=R(x,y,z)\, \left({\over}\alpha\, e^{{i\over \hbar}
S_0(\vec{r})} +\beta e^{-{i\over \hbar} S_0(\vec{r})} \right) \; ,
\end{equation}
so that, the last expression (\ref{wavefunction1}) of the wave function is also a
solution of each one of the Eqs. (\ref{1d-schro-x}). It follows that the
functions $S_0$ and $R$ satisfy the following equations
\cite{Dja4,Dja6}
\begin{eqnarray}\label{1d-qshje-x}
{1 \over 2m}\left({\partial S_0 \over \partial
x_i}\right)^2-{\hbar^2 \over 2m}{1\over R}{\partial^2 R\over
\partial x_i^2}+V_{x_i}(x_i)=E_{x_i} \; ,\\ \label{1d-current-x} {\partial
\over \partial x_i}\left(R^2 {\partial S_0 \over \partial
x_i}\right)=0 \; ,\hskip40mm\\ i=1,2,3.\hskip10mm  x_1=x, x_2=y, x_3=z\hskip10mm\nonumber
\end{eqnarray}
These last equations are easily obtained by replacing expression
(\ref{wavefunction1}) into Eqs. (\ref{1d-schro-x}) exactly as
used to be obtained in the literature of the Bohmian quantum
mechanics \cite{Floyd1, Bohm1, Farmat1}. The integration of
Eqs. (\ref{1d-current-x}) is obvious and leads to
\begin{eqnarray}\label{amplitude2}
\label{1d-R-x} R(x,y,z)=f(y,z) \left({\partial S_0 \over \partial
x}\right)^{-{1\over 2}} \; ,\hskip35mm\\ \label{1d-R-y}
R(x,y,z)=g(x,z) \left({\partial S_0 \over \partial
y}\right)^{-{1\over 2}} \; ,\hskip35mm\\ \label{1d-R-z}
R(x,y,z)=k(x,y) \left({\partial S_0 \over \partial
z}\right)^{-{1\over 2}} \; .\hskip35mm
\end{eqnarray}
Now, taking Eqs. (\ref{1d-R-x}), (\ref{1d-R-y}) and (\ref{1d-R-z})
into Eqs. (\ref{1d-qshje-x}), we find
\begin{eqnarray}\label{1d-qshje2-x}
{1 \over 2m}\left({\partial S_0 \over \partial
x_i}\right)^2-{\hbar^2 \over
4m}\left\{{\over}S_0,x_i\right\}+V_{x_i}(x_i)=E_{x_i} \; ,\\
i=1,2,3.\hskip8mm  x_1=x, x_2=y, x_3=z\, .\hskip10mm\nonumber
\end{eqnarray}
where $\left\{{\over}S_0,q\right\}$ is the Schwartzian derivative
of $S_0$ with respect to $q$. Thus, we obtained three 1D-QSHJE
instead of one 3D-QSHJE and three 1D current's conservation equations
instead of one 3D current's conservation equation.

\noindent Even though the reduced action given in Eq. (\ref{sep-action}) is solution of Eqs. (\ref{1d-qshje2-x}),
it is not the only possible solution, and as it is stated in Ref. \cite{boud1}, another
possible solution of the 3D-QSHJE is of the form
\begin{equation}\label{3d-action-nonsep}
S_{0}^{(2)}(x,y,z)=\hbar \arctan{\left({\sum_{i, j,k =1}^{2}
a_{ijk}X_{i} Y_j Z_k\over \sum_{i, j,k =1}^{2}
b_{ijk}X_{i} Y_j Z_k}\right)}
\end{equation}
up to an additive constant. Note that $S_{0}^{(2)}$ is also
solution of Eqs. (\ref{1d-qshje2-x}).

\noindent Notice that Eqs. (\ref{1d-qshje2-x}) are of the second order
in $\partial S_0/ \partial x_i$ with respect to $x_i$. That means
after integrating Eqs. (\ref{1d-qshje2-x}), the quantum conjugate
momenta must have, each one, two integration constants, giving a total
of six different integration constants. Thus, the total reduced action
has to have nine integration constants, three of them are just additive
constants that can be reduced to be one additive constant.

In Ref. \cite{boud2}
it is stated and shown that the form given in Eq.
(\ref{sep-action}) is a particular case of Eq.
(\ref{3d-action-nonsep}). We totally agree with that. Furthermore, for the authors of \cite{boud2}
there are fifteen integration
constants completely independents, one of them is an additive one. In Ref.
\cite{boud2}, it was shown that if one considers a solution of the
form given in Eq. (\ref{3d-action-nonsep}), excluding the additive constant,
not all of the sixteen
constants do play the role of integration constants and two of
them have to be fixed. For the authors of Ref. \cite{boud2}
$a_{111}=1$ and $b_{222}=1$.

\noindent For us, both forms given in Eqs. (\ref{sep-action}) and
(\ref{3d-action-nonsep}) are solutions of the 3D-QSHJE in symmetrical potential case, however for us, only expression (\ref{sep-action}) is a physical solution \footnote{$S_0^{(2)}$ contains unnecessary redundant constants, the fourteen constants are not free each one from the others. In addition, when trying to find the quantum trajectories, we should have a trajectory for each combination of the integration constants. With a set of fourteen constants, we will have largely more possible combinations than with a set of six constants. So, if the total number of the integration constants is fourteen, then each combination will describe a real possible quantum trajectory of the particle. However, if the total number of the integration constants is just six (up to an additive constant), then, if one uses $S_0^{(2)}$, there will be a big number of combinations of the fourteen constants that will describe non-real trajectories that does not exist. Furthermore, $S_0^{(1)}$ contains only six constants, and each combination of the set of these six constants will describe a physical quantum trajectory. So, for us $S_0^{(1)}$ is a physical solution while $S_0^{(2)}$ is not a physical one.} and contains the valid integration constants. We believe strongly that the number of these integration constants is six plus a seventh additive one. That is what we will show in what follows. The reason of our belief is that the 3D-QSHJE is separable into three 1D-QSHJE, while for Bouda and al this 3D-QSHJE is not separable when they deal with symmetrical potential cases.
In order that someone identifies the integration constants, he has first to integrate Eqs. (\ref{1d-qshje2-x}). Then, let us start with Eq. (\ref{1d-qshje2-x}) written with respect to $x$. To proceed, we will follow the procedure of Faraggi and Matone \cite{Farmat3}. Faraggi and Matone showed that the following equation
\begin{equation}\label{equation1}
{\partial^2 \psi \over \partial x^2}- {1\over 2}\left\{{\over}F ,x\right\}\psi\, =\, 0 \end{equation}
has a solution of the form
\begin{equation}\label{equation2}
\psi \, =\, \left(\partial F \over \partial x\right)^{-{1\over2}}\left({\over}a(y,z) F +b(y,z)\right)\, . \end{equation}
a, and b are constants with respect to $x$, but functions of $y$ and $z$.
If we choose the function $F$ such as
$$
F(x,y,z) \, =\, e^{{2 i\over \kappa}S_0} \, ,
$$
then, Eq. (\ref{equation1}) will correspond to the SE after identifying $\kappa$ to $\hbar$ \cite{Farmat3}.

\noindent The solution of the 1D-QSHJE (\ref{1d-qshje2-x})
is related to the two solutions (Eq. (\ref{equation2})) given by
\begin{equation}\label{equation6}
\Psi_1(x,y,z) = -\hbar^2\, \left({\partial S_0\over \partial x}\right)^{-{1\over 2}}\, e^{-{i\over \hbar}S_0}\, \left({\over}a(y,z)\, e^{{2i\over \hbar}S_0}+b(y,z)\right)
\end{equation}
and
\begin{equation}\label{equation7}
\Psi_2(x,y,z) = -\hbar^2\, \left({\partial S_0\over \partial x}\right)^{-{1\over 2}}\, e^{-{i\over \hbar}S_0}\, \left({\over}c(y,z)\, e^{{2i\over \hbar}S_0}+d(y,z)\right)
\end{equation}
where $a$, $b$, $c$ and $d$ are complex functions.
Remark that the above solutions $\Psi_1$ and $\Psi_2$ are functions of the three
variables $x$, $y$ and $z$, so that they are identical to expression (\ref{wavefunction1}) of the wave function, $\Psi_1$ and $\Psi_2$ has also the form given in Eq. (\ref{psitot1}). Then, if one compares Eqs. (\ref{equation6}) and (\ref{equation7}) to Eq. (\ref{wavefunction1}) and taking account of Eq. (\ref{amplitude2}), these two last equations can be written as
$$
\Psi_1 = -\hbar^2\, \left({\partial S_0\over \partial x}\right)^{-{1\over 2}}\, f(y,z)\, \left({\over}a^{'}\, e^{{i\over \hbar}S_0}+b^{'}e^{-{i\over \hbar}S_0}\right)
$$
and
$$
\Psi_2 = -\hbar^2\, \left({\partial S_0\over \partial x}\right)^{-{1\over 2}}\, f(y,z)\, \left({\over}c^{'}\, e^{{i\over \hbar}S_0}+d^{'}e^{-{i\over \hbar}S_0}\right)
$$
where $a^{'}$, $b^{'}$, $c^{'}$ and $d^{'}$ are complex constants.
Finally, we can write
\begin{equation}\label{equation9}
S_0(x,y,z)\, =\, {\hbar \over 2i} \ln \left[{-d^{'} \Psi_1 + b^{'} \Psi_2\over c^{'} \Psi_1 - a^{'}\Psi_2}\right]
\end{equation}
In fact this last solution of the 1D-QSHJE (\ref{1d-qshje2-x}) must contain only seven integration constants. This is why we shall write
\begin{eqnarray}\label{equation11}
 S_0(x,y,z)\, =\,
 \hbar \arctan\left[{\Psi^{'}_1(x,y,z) \over  \Psi^{'}_2(x,y,z)}\right]\, + \hbar \lambda_0,\hskip30mm\nonumber\\
 =\, \hbar \arctan\left[{\sum_{i, j,k =1}^{2}
a^{'}_{ijk}X_{i} Y_j Z_k\over \sum_{i, j,k =1}^{2}
b^{'}_{ijk}X_{i} Y_j Z_k}\right]\, + \hbar \lambda_0\, ,\hskip15mm
\end{eqnarray}
where $\lambda_0$ represent the additive integration constant.
With the same way we get after solving the two other equations
in Eqs. (\ref{1d-qshje2-x}) to Eq. (\ref{equation11}).
Notice that the form of the solution looks like the expression
given by Eq. (\ref{3d-action-nonsep}). However, since there are
only six integration constants inside the arc tangent argument,
the sixteen constants $a^{'}_{ijk}$ and $b^{'}_{ijk}$ must be
function of the six integration constants. In this case the form 
of the solution must reduce to the expression $S^{(1)}_0$ given
by Eq. (\ref{sep-action}). This is has already been shown in Ref. \cite{boud2}.
Notice that we admitted implicitly that the integration constants are pure constants.
This might be a subject of a new criticism since Eqs.(\ref{1d-qshje2-x})
are partial differential equations and not ordinary ones. Indeed, one
might consider that the three integration constants that come out from
the first of Eqs. (\ref{1d-qshje2-x}) depend explicitly on $y$, and $z$.
The three integration constants that come out from the second of
Eqs. (\ref{1d-qshje2-x}) depend explicitly on $x$, and $z$. And, the
three integration constants that come out from the third of
Eqs. (\ref{1d-qshje2-x}) depend explicitly on $x$, and $y$.
So that Eqs. (\ref{1d-qshje2-x}) integrate to give
\begin{equation}\label{equation11a}
 S_0(x,y,z)\, =\, \hbar \arctan\left[{X_1(x) + \Gamma_1(y,z) X_2(x)\over \Gamma_2(y,z) X_1(x)+X_2(x)}\right]\, + \hbar \lambda_1(y,z),
\end{equation}
\begin{equation}\label{equation12a}
 S_0(x,y,z)\, =\, \hbar \arctan\left[{Y_1(y) + \Gamma_3(x,z) Y_2(y)\over \Gamma_4(x,z) Y_1(y)+Y_2(y)}\right]\, + \hbar \lambda_2(x,z),
\end{equation}
and
\begin{equation}\label{equation13a}
 S_0(x,y,z)\, =\, \hbar \arctan\left[{Z_1(z) + \Gamma_5(x,y) Z_2(z)\over \Gamma_6(x,y) Z_1(z)+Z_2(z)}\right]\, + \hbar \lambda_3(x,y)\, .
\end{equation}
respectively. If we consider that the solution of the 3D-QSHJE is of the form (\ref{3d-action-nonsep})
and that this form cannot be written as the sum of three 1D reduced actions \cite{boud2}, we must have
$$
\lambda_1(y,z)=\lambda_2(x,z)=\lambda_3(x,y)=\lambda\, .
$$
where $\lambda$ is the additive integration constant.
Since the three forms of the reduced action given in Eqs. (\ref{equation11a}), (\ref{equation12a})
and (\ref{equation13a}) must be equals, we must have
$$
{X_1(x) + \Gamma_1(y,z) X_2(x)\over \Gamma_2(y,z) X_1(x)+X_2(x)}={Y_1(y) + \Gamma_3(x,z) Y_2(y)\over \Gamma_4(x,z) Y_1(y)+Y_2(y)}={Z_1(z) + \Gamma_5(x,y) Z_2(z)\over \Gamma_6(x,y) Z_1(z)+Z_2(z)}\, ,
$$
In order to reproduce the form $S_0^{(2)}$
given in Eq. (\ref{3d-action-nonsep}) starting froms (\ref{equation11a}-\ref{equation13a}). we can set
$$
 \Gamma_1(y,z)\, =\, \Gamma_1^{11} Y_1(y) Z_1(z)+\, \Gamma_1^{12} Y_1(y) Z_2(z)+\, \Gamma_1^{21} Y_2(y) Z_1(z)+\, \Gamma_1^{22} Y_2(y) Z_2(z),
$$
$$
 \Gamma_2(y,z)\, =\, \Gamma_2^{11} Y_1(y) Z_1(z)+\, \Gamma_2^{12} Y_1(y) Z_2(z)+\, \Gamma_2^{21} Y_2(y) Z_1(z)+\, \Gamma_2^{22} Y_2(y) Z_2(z),
$$
$$
 \Gamma_3(x,z)\, =\, \Gamma_3^{11} X_1(x) Z_1(z)+\, \Gamma_3^{12} X_1(x) Z_2(z)+\, \Gamma_3^{21} X_2(x) Z_1(z)+\, \Gamma_3^{22} X_2(x) Z_2(z),
$$
$$
 \Gamma_4(x,z)\, =\, \Gamma_4^{11} X_1(x) Z_1(z)+\, \Gamma_4^{12} X_1(x) Z_2(z)+\, \Gamma_4^{21} X_2(x) Z_1(z)+\, \Gamma_4^{22} X_2(x) Z_2(z),
$$
If we take these last expressions of $\Gamma_i$ into Eqs. (\ref{equation11a}), (\ref{equation12a})and (\ref{equation13a}), we notice that the quotient inside the arctangent contains ten terms, while in $S_0^{(2)}$ it  contains sixteen terms. In addition, the terms $X_1$ and $X_2$ present in Eq. (\ref{equation11a}) are not present in the two other equations, the terms $Y_1$ and $Y_2$ present in Eq. (\ref{equation12a}) are not present in the two other equations, the terms $Z_1$ and $Z_2$ present in Eq. (\ref{equation13a}) are not present in the two other equations.
This means that it is not possible to get the form $S_0^{(2)}$ from the integrated reduced action given in Eqs. (\ref{equation11a}), (\ref{equation12a})and (\ref{equation13a}). Furthermore, in this case, the reduced actions given in
Eqs. (\ref{equation11a}), (\ref{equation12a})and (\ref{equation13a}) cannot be equal, which is in contradiction with the above hypotheses. So in Eqs. (\ref{equation11a}), (\ref{equation12a}) and , (\ref{equation13a}) the six functions $\Gamma_i$ must be replaced by the constants $\gamma_i$, and the three functions $\lambda_i$ must depend on $x$, $y$ and $z$.
At the first sight, the function $\lambda_1$ should be equal to
the sum of the two artangents present in the other two equations.
This can be shown by taking the derivative of Eq. (\ref{equation11a}) three times with
respect to $y$ and replacing them into the second of Eqs. (\ref{1d-qshje2-x}), after which we
find that $\lambda_1(y,z)$ satisfies the QSHJE. It means that
$$
\lambda_1(y,z)=\, \hbar \arctan\left[{Y_1(y) + \gamma_3 Y_2(y)\over \gamma_4 Y_1(y)+Y_2(y)}\right]\, + \hbar {\cal B}(z)\, .
$$
If we take the last equation into Eq. (\ref{equation11a}) and take the derivative of this last three times with respect to $z$ and replace them into the last of Eqs. (\ref{1d-qshje2-x}), we will find that ${\cal B}(z)$ satisfies
the QSHJE. It means that
$$
{\cal B}(z)=\, \hbar \arctan\left[{Z_1(z) + \gamma_5 Z_2(z)\over \gamma_6 Z_1(z)+Z_2(z)}\right]\, + \hbar\lambda \, .
$$
In the end we reach the result that the 3D reduced action must be written as expressed in Eq. (2).

\noindent Now, let us talk about the amplitude of the wave function $R(x,y,z)$.
How can we write it? First of all, let us inject expression (\ref{1d-R-x})
into Eq. (\ref{1d-current-x}) written with respect to $y$, we get
\begin{equation}\label{ampl1}
{\partial \over \partial y}\left[f^2(y,z)\left({\partial S_0 \over \partial x} \right)^{-1}\left({\partial S_0 \over \partial y} \right)\right]=\, 0\, ,
\end{equation}
but $\partial S_0 /\partial x$ depends only on $x$ and can be put outside the derivative with respect to $y$. Thus, Eq. (\ref{ampl1}) reduces to
$$
{\partial \over \partial y}\left[f^2(y,z)\left({\partial S_0 \over \partial y} \right)\right]\, =\, 0\, ,
$$
then the last equation integrates to give
$$
f(y,z)= {\cal{N}}(z) \left({\partial S_0 \over \partial y} \right)^{-{1\over 2}} \, ,
$$
and we can write for the amplitude $R(x,y,z)$ the following expression
$$
R(x,y,z)= {\cal{N}}(z) \left({\partial S_0 \over \partial x} \right)^{-{1\over 2}}\left({\partial S_0 \over \partial y} \right)^{-{1\over 2}} \, .
$$
Now, let us take the last expression of the amplitude into Eq. (\ref{1d-current-x}) written with respect to $z$, we will find
$$
{\partial \over \partial z}\left[{\cal{N}}^2(z)\left({\partial S_0 \over \partial z} \right)\right]\, =\, 0\, ,
$$
integrating this equation we get to
$$
{\cal{N}}(z)=\, k\; \left({\partial S_0 \over \partial z} \right)^{-{1\over 2}}\, ,
$$
where k is a real constant. Thus, the final expression of the amplitude of the wave function is
$$
R(x,y,z)= k\; \left({\partial S_0 \over \partial x} \right)^{-{1\over 2}} \left({\partial S_0 \over \partial y} \right)^{-{1\over 2}}\left({\partial S_0 \over \partial z} \right)^{-{1\over 2}} \, .
$$
In fact, we get to a separable form of the amplitude since all the derivatives with respect to $x$, $y$ and $z$ depend only and separately on the variables $x$, $y$ and $z$ respectively. Let us write
\begin{eqnarray}\label{Sep-ampl-x}
R_{x_i}(x_i)= k_{x_i}\; \left({\partial S_0 \over \partial x_i} \right)^{-{1\over 2}}\, ,
\hskip40mm\nonumber\\
i=1,2,3.\hskip8mm  x_1=x, x_2=y, x_3=z\, .\hskip17mm\nonumber
\end{eqnarray}
Thus
$$
R(x,y,z)= R_x(x) R_y(y) R_z(z)\, .
$$
We stress that we ended by coming back to the initial hypotheses advanced in Refs. \cite{Dja4,Dja5,Dja6}, concerning the separability of the reduced action and the amplitude of the wave function.

\vskip0.5\baselineskip

Now, let us look for how to express the dynamical laws of
motion in the case of a potential with a cartesian symmetry.
Firstly, starting from Eqs.(\ref{1d-qshje2-x}), we can write \cite{Farmat3}
\begin{eqnarray}\label{1d-energy-x}
{P_{x_i}^2 \over 2m}\, g_{x_ix_i}(x_i)+V_{x_i}(x_i)=E_{x_i} \; ,\hskip30mm\\
i=1,2,3.\hskip8mm  x_1=x, x_2=y, x_3=z\, .\hskip17mm\nonumber\; ,
\end{eqnarray}
where
\begin{equation}\label{1d-mom-x}
P_{x_i}={\partial S_0 \over \partial x_i}\; ,
\end{equation}
and
\begin{equation}\label{1d-metric1-x}
g_{x_ix_i}(x_i)=\left[1-{\hbar^2 \over 2} \left({\partial S_0
\over \partial x_i}\right)^{-2}\left\{{\over}S_0,x_i\right\}\right] \; .
\end{equation}
This approach has already been introduced by Faraggi and Matone
through the quantum coordinate \cite{Farmat3}.
Now, if one sums the three Eqs. (\ref{1d-energy-x}), he finds
\begin{equation}\label{1d-totenergy-x}
\sum_{i=1}^{3} {P_{x_i}^2 \over 2m}\, g_{x_ix_i}(x_i)+V(x,y,z)=E \;
.
\end{equation}
In the last equation, if we identify the energy to the hamiltonian
of the quantum system and using the Hamilton's canonical
equations $\dot{x}_i=\partial H/\partial P_{x_i}$, we will get to
\begin{equation}\label{1d-mom-metric-x}
P_{x_i}\, g_{x_ix_i}=m\dot{x}_i\; .
\end{equation}
Taking Eqs. (\ref{1d-mom-metric-x})into Eqs. (\ref{1d-energy-x}), we get
\begin{eqnarray}\label{1d-qlm-x}
\dot{{x}_i}\, {\partial S_0 \over \partial x_i}=2[E_{x_i}-V_{x_i}(x_i)] \; ,\hskip35mm\\
i=1,2,3.\hskip8mm  x_1=x, x_2=y, x_3=z\, .\hskip17mm\nonumber\; .
\end{eqnarray}
These last three equations represent the 3D quantum law of motion in the case
of a cartesian symmetry potential. Let us recall that a familiar form of
equation of motion has already been investigated by Goldstein \cite{Gold}
in the classical case. However, in Eqs. (\ref{1d-qlm-x}), the quantum nature
of the equations is apparent. Indeed, these equations hide more information
than its corresponding classical one since it contains the six integration constants
$\gamma_i$. So, Eqs. (\ref{1d-qlm-x}) describes many quantum trajectories while
their corresponding classical ones describes a unique trajectory for a
fixed set of initial conditions.

Notice that if one considers $S_0^{(2)}$ as the reduced action of
the system, then Eqs. (\ref{1d-qlm-x}) will contain each one fifteen integration
constants, the fourteen constants already present in $\partial
S_0^{(2)}/\partial x_i$ and the energies $E_{x_i}$ such as
the number of constants is seventeen. So, in order to eliminate
these constants in all the three equations, one must take the
derivative with respect to time fifteen times, which leads to
equations of motion of the sixteenth degree with respect to time.
Due to such complexity that generates a reduced action of the form
$S_0^{(2)}$, we were skeptical to consider it as the physical
solution of the 3D-QSHJE for the case of separable variables.
Nevertheless, our skepticism does not allow us to exclude
$S_0^{(2)}$. In the opposite,
if we use the reduced action $S_0^{(1)}$, Eqs. (\ref{1d-qlm-x}) will contain, each one, two integration
constants in addition to the energies $E_x$, $E_y$ and $E_z$. This means that
after taking the derivative three times with respect to time, we can get rid of
these integration constants and have equations of the fourth order with respect to time.

Here, it is useful to clarify one point.
It concerns the dynamical Eqs. (\ref{1d-mom-metric-x}) and the identities given
in Eqs. (\ref{1d-metric1-x}). In fact, we can see clearly in  Eqs. (\ref{1d-metric1-x})
that $g_{x_ix_i}$ depend explicitly on $P_{x_i}$, $\partial P_{x_i}/ \partial x_i$
and $\partial^2 P_{x_i} /\partial x_i^2$. So, applying the Hamilton's canonical
equations would lead to
\begin{equation}\label{1d-mom-metric-xbis}
m\dot{x}_i=P_{x_i}\, g_{x_ix_i}+{P_{x_i}^2\over 2}{\partial g_{x_ix_i}\over \partial P_{x_i}}\; .
\end{equation}
which is different from Eqs. (\ref{1d-mom-metric-x}). So which one of the
two Eqs. (\ref{1d-mom-metric-x}) and (\ref{1d-mom-metric-xbis}) is correct.
This is a very controversial point and it raises
contradiction in the quantum equations of motion. For example, if one writes the 1D classical Hamiltonian as
$$
H={1\over 2}\, P\dot{x}+V(x)
$$
where, we identified $P$ to $m\dot{x}$, we get to a contradiction
($\dot{x}/2=\dot{x}$) when one applies the canonical equation
$\partial H/\partial P=\dot{x}$.
To explain this point, first of all notice that $g_{x_ix_i}$ indicate a deformation of the quantum space,
so we should link this deformation of the geometry to spatial coordinates $x_i$ rather than to momenta
of motion $P_{x_i}$. The link between the quantum momenta of motion and the spatial coordinates that appears
in Eqs. (\ref{1d-metric1-x}) will be established after identifying the equations of motion.
Indeed, when Faraggi and Matone derived the QSHJE by using the quantum equivalent postulate \cite{Farmat1,Farmat2,Farmat3}, they started from the Legendre transformation of the Hamilton's characteristic function
$$
S_0(q)=Pq-{\cal T}_0(P)
$$
$$
P={\partial S_0\over\partial q},\hskip10mm q={\partial {\cal T}_0\over\partial P}
$$
${\cal T}_0$ being the dual of $S_0$.
Choosing $S_0(q,P(q))$ in the Legendre transformation
means that Faraggi and Matone admitted implicitly that the
quantum Hamiltonian is a function of just $q$ and $P$,
since $\partial S_0(q,P(q),t)/ \partial t=-H(q,P,t)$.

\noindent Faraggi and Matone wrote the QSHJE as
$$
{1\over 2m}\left({\partial S_0\over \partial q}\right)^2+Q(q)+V(q)=E\, ,
$$
where $Q(q)$ is the quantum potential that is set to be dependent on $q$.
They ended by identifying the quantum potential to be
$$
Q(q)=-{\hbar^2\over 4m}\{S_0(q),q\}\, ,
$$
which contains $P$, $\partial P/\partial q$ and $\partial^2 P/\partial q^2$, making the QSHJE depending on these quantities. The QSHJE being an equation of motion, $q$ is linked now to $P$, $\partial P/\partial q$ and $\partial^2 P/\partial q^2$ through the expression of $Q(q)$ given above. Note that, if one has to consider $Q$ depending on $P$ and its derivatives in the QSHJE and in the quantum Hamiltonian, at the first approach, he has first to consider a new Legendre transformation that must contain not only $q$ and $P$, but also $\partial P/\partial q$ and $\partial^2 P/\partial q^2$. The same reflection can be considered for the Hamiltonian (\ref{1d-totenergy-x}) and $g_{x_ix_i}(x_i)$. In fact, this Hamiltonian depends only on $x_i$ and $P_{x_i}$, so its corresponding Legendre transformation contains only $x_i$ and $P_{x_i}$. However, if one considers that the quantities $g_{x_ix_i}$ depend on $P_{x_i}$, $\partial P_{x_i}/\partial x_i$ and $\partial^2 P_{x_i}/\partial x_i^2$, the Hamiltonian of the system will depend on the derivatives of $P$ so its corresponding Legendre transformations must contain $x_i$, $P_{x_i}$, $\partial P_{x_i}/\partial x_i$ and $\partial^2 P_{x_i}/\partial x_i^2$, leading to a different QSHJE than the one we know. Finally, we stress that $g_{x_ix_i}$ appearing in the expression of the quantum Hamiltonian must depend explicitly on $x_i$, and the correct expression between Eqs. (\ref{1d-mom-metric-x}) and (\ref{1d-mom-metric-xbis}) is Eqs. (\ref{1d-mom-metric-x}).

Now, let us consider Eqs. (\ref{1d-qlm-x})
in the neighborhood of the classical turning points. It is clear from these
equations that in the classically allowed regions where $V_{x_i}(x_i)<E_{x_i}$,
$\dot{{x}_i}$ and $\partial S_0 /\partial x_i$ must have same signs,
meanwhile in the classically forbidden regions $V_{x_i}(x_i)>E_{x_i}$
$\dot{{x}_i}$ and $\partial S_0 /\partial x_i$ must have opposite signs.
This suggest that $\partial S_0 /\partial x_i$ must have opposite signs on different
sides of a classical turning point. Thus, one may think that at a classical turning point where
$V_{x_i}(x_i)=E_{x_i}$, the quantum momenta $\partial S_0 /\partial x_i$ takes
a zero value contradicting the finding of Faraggi and Matone \cite{Farmat3}
which states that $\partial S_0 /\partial x_i$ can never be nil for a finite $x_i$.
Indeed, from Eq. (\ref{sep-action})
\begin{equation}\label{quantmompm}
{\partial S_0 \over \partial x}=\pm{\hbar (1-\gamma_1\gamma_2)W_x \over (X_1+\gamma_1 X_2)^2+(\gamma_2 X_1+ X_2)^2}, ,
\end{equation}
where $W_x$ is the Wronskian of $X_1$ and $X_2$. The $\pm$ sign indicates
that the motion can be in either direction on the $x$ axis. This sign is closely related to progressive
waves $\exp(+i{S_0\over \hbar})$ and $\exp(-i{S_0\over \hbar})$.
If $\partial S_0 / \partial x=0$, then $1-\gamma_1\gamma_2=0$ which makes $S_0$ a constant,
or $W_x=0$ which makes the two solutions $X_1$ and $X_2$ dependent, and makes $S_0$ a constant.
Therefore, there is no way for $\partial S_0 / \partial x$ to be nil. The same conclusion is correct for
$\partial S_0 / \partial y$ and $\partial S_0 / \partial z$. So, what would happen in the vicinity of a classical turning point? In our opinion, when a particle approaches a classical turning point in the classically allowed side, it start to loose its kinetic energy $E_{x_i}-V_{x_i}(x_i)$ until it reaches the turning point ($E_{x_i}=V_{x_i}(x_i)$) where $\dot{x}_i^{tp}=0$ instantly and locally. There, the particle has two possibilities, either it comes back in the classically allowed regions, so that $\partial S_0 / \partial x_i$ keeps the same sign as $\dot{x}_i$, or the particle will enter the classically forbidden regions, so that $\partial S_0 / \partial x_i$ changes the sign. This change happens instantly and with discontinuity. The discontinuity of
$\partial S_0 / \partial x_i$ occurs through it expression (\ref{quantmompm}) by changing it sign when the particles crosses the turning point wether it goes in the classically forbidden region or it comes out from it. Indeed, in the classically allowed regions, since $E_{x_i}>V_{x_i}$, $\dot{x}_i$ and $\partial S_0 / \partial x_i$ have both positive sign when the motion is toward the positive $x_i$ and negative when the motion is toward the negative $x_i$. Then, for the classically allowed regions, the part of the wave carrying the motion toward the positive $x_i$ is $\exp(+i{S_0\over \hbar})$, while the one carrying the motion toward the negative $x_i$ is $\exp(-i{S_0\over \hbar})$. This is in accordance with the results of the standard quantum mechanics. In the classically forbidden regions, since $E_{x_i}<V_{x_i}$, $\dot{x}_i$ and $\partial S_0 / \partial x_i$ have opposite signs, when the motion is toward the positive $x_i$, $\partial S_0 / \partial x_i$ is negative and the part of the wave carrying the motion is $\exp(-i{S_0\over \hbar})$, however, when the motion is toward the negative $x_i$, $\partial S_0 / \partial x_i$ is positive and the part of the wave carrying the motion is $\exp(+i{S_0\over \hbar})$. At the first sight, this last observation maybe though to be in contradiction with common sense, in other words, how would $\exp(+i{S_0\over \hbar})$ carry the motion toward the negative $x_i$ and $\exp(-i{S_0\over \hbar})$ carry the motion toward the positive $x_i$. In fact, when the particle enters the classically forbidden regions and moves toward the positive $x_i$, the forbidden region tries to push it back outside, letting $\exp(-i{S_0\over \hbar})$ be the carrier of the particle's motion, and when the particle moves toward the negative $x_i$, the forbidden region opposes its motion and tries to keep it inside it by letting $\exp(+i{S_0\over \hbar})$ be the carrier of the motion. This makes the dwell time inside the forbidden regions  finite even if these last ones are thick which is in accordance with previous works \cite{Dja2}.

\vskip7mm

To conclude, we can say that we successfully established
the separability of the 3D-QSHJE
into three 1D-QSHJE. This separability implies expressing
the total reduced action as the sum of the three 1D reduced
actions. This separability of the 3D-QSHJE and the reduced action
is valid for any kind of symmetrical potential case, cartesian,
cylindrical or even spherical symmetry potentials and also for some separated variables potentials.

\vskip3\baselineskip \noindent {\bf REFERENCES}

\end{document}